\documentclass{article}
\usepackage[breaklinks]{hyperref}
\usepackage{fancyheadings}
\usepackage[german,british]{babel}

\usepackage{amsmath}
\usepackage{amssymb}

\usepackage{graphicx}
\usepackage{psfig}
\usepackage{color}

\newcommand{\dif}{\mathrm{d}}

\DeclareFontFamily{OT1}{rsfs}{}
\DeclareFontShape{OT1}{rsfs}{m}{n}{ <-7> rsfs5 <7-10> rsfs7 <10->rsfs10}{} 
\DeclareMathAlphabet{\mycal}{OT1}{rsfs}{m}{n}
\newcommand{\scri}{{\mycal I}}

\newcommand{\arccosh}{\mathrm{arccosh}}

\numberwithin{equation}{section}

\begin{document} 

\pagenumbering{arabic}
\title{Eleven spherically symmetric constant density solutions with cosmological constant}
\author{Christian G. B\"ohmer\footnote{E-mail: boehmer@hep.itp.tuwien.ac.at}}
\date{}
\maketitle
\thispagestyle{fancy}
\setlength{\headrulewidth}{0pt}
\rhead{}
\rhead{TUW--03--37}
\noindent
\textit{Institut f\"ur Theoretische Physik, Technische Universit\"at Wien, 
Wiedner Hauptstrasse 8-10, A-1040 Wien, \"Osterreich.}
\\ \mbox{} \\
\begin{abstract}
Einstein's field equations with cosmological constant are analysed 
for a static, spherically symmetric perfect fluid having constant density.
Five new global solutions are described. 

One of these solutions has the Nariai solution joined on as 
an exterior field. Another solution describes a decreasing pressure 
model with exterior Schwarzschild-de Sitter spacetime having 
decreasing group orbits at the boundary. 
Two further types generalise the Einstein static universe.

The other new solution is unphysical, it is an increasing pressure 
model with a geometric singularity.
\end{abstract}
\mbox{} \\ 
\mbox{} 

\newpage
\section{Introduction}
Cosmological observations~\cite{Krauss:2003yb,Perlmutter:1998zf} give
strong indications for the presence of a positive cosmological constant,
on the other hand the low energy limit of supersymmetry theories
requires a negative sign~\cite{VanNieuwenhuizen:1981ae}.

Therefore it is interesting to analyse solutions to the field equations
with cosmological constant representing for example relativistic stars.
An overview of these solutions is given in table~\ref{oview}, 
appendix~\ref{app_over}.

With vanishing cosmological constant the first static and spherically 
symmetric perfect fluid solution with constant density was already found 
by Schwarzschild in 1916~\cite{Schwarzschild:1916ae}.
In spherical symmetry Tolman~\cite{Tolman:1939} and Oppenheimer 
and Volkoff~\cite{Oppenheimer:1939ne} reduced the field equations
to the well known TOV equation.
For simple equations of state Tolman integrated the TOV equation 
and discussed solutions. Although he included the cosmological 
constant in his calculations he stated that it is 
too small to produce effects. The geometrical properties of
constant density were analysed by Stephani~\cite{Stephani:1967}.

With non-vanishing cosmological constant $\Lambda\neq0$ constant density 
solutions were first analysed by Weyl~\cite{Weyl:1919}. 
The different possible spatial 
geometries were already pointed out and the possible coordinate 
singularity was mentioned. 
More than 80 years later Stuchl{\'{\i}}k~\cite{Stuchlik:2000} analysed
these solutions again. He integrated the 
TOV-$\Lambda$ equation for possible values of the cosmological
constant up to the limit \mbox{$\Lambda < 4\pi \rho_{0}$},
where $\rho_{0}$ denotes this constant density.
In these cases constant density solutions describe stellar models. 
In the present paper it is shown that a coordinate singularity occurs
if $\Lambda \geq 4\pi \rho_{0}$, already mentioned by Weyl~\cite{Weyl:1919}.

If the cosmological constant equals this upper bound the pressure vanishes 
at the mentioned coordinate singularity. In this case one has to use the 
Nariai solution \cite{Nariai:1999nc,Nariai:1999os} 
to get the metric $C^{1}$ at the boundary. 
For larger cosmological constant the pressure will vanish after 
the coordinate singularity. The volume of group orbits is decreasing 
and there one has to join the Schwarzschild-de Sitter solution 
containing the $r=0$ singularity. Increasing the cosmological 
constant further leads to generalisations of the Einstein static 
universe. These solutions have two regular centres with 
monotonically decreasing or increasing pressure from the
first to the second centre. Certainly the Einstein cosmos itself is a 
solution.

Solutions with $\Lambda\geq 4\pi\rho_{0}$ have not been analysed in
literature so far. Thus the described constant density solutions 
obeying this bound are considered to be new, except for the 
Einstein static universe. 
\section{Cosmological TOV equation}
\label{costov}

Consider the following line element
\begin{align}
      \dif s^{2}=-\frac{1}{\lambda} e^{\lambda \nu (r)}\dif t^{2}+
      e^{a(r)}\dif r^2+r^{2}\dif\Omega^{2},
      \label{sssmetric}
\end{align} 
where $G=1$ and $\kappa=8\pi \lambda^{2}$ will be used. 
$\lambda=1$ corresponds to Einstein's theory of gravitation 
in geometrised units, roughly speaking 
$\lambda \rightarrow 0$ corresponds to the Newtonian 
theory~\cite{Ehlers:1989,Ehlers:1991,Ehlers:1997}.

The field equations for a perfect fluid are three independent equations, 
which imply energy-momentum conservation. One may either use the three 
independent field equations or one uses two field equations and the 
energy-momentum conservation equation. For the purpose 
of this paper it is more convenient to do the second. 
Hence consider the first two field equations and the 
conservation equation
\begin{align}
      \frac{1}{\lambda r^{2}}e^{\lambda \nu(r)}\frac{\dif}{\dif r}\left(r-re^{-a(r)}\right) 
      -\Lambda e^{\lambda \nu(r)} &= 8\pi \rho(r)e^{\lambda \nu(r)}, 
      \label{a} \\
      \frac{1}{r^{2}}\left(1+r\lambda \nu'(r)-e^{a(r)}\right) 
      +\lambda\Lambda e^{a(r)} &= 8\pi \lambda^{2} P(r)e^{a(r)},
      \label{b} \\
      -\frac{\nu'(r)}{2}(\lambda P(r)+\rho) &= P'(r).
      \label{c} 
\end{align}
There are four unknown functions. From a physical point of view there are two 
possibilities. Either a matter distribution $\rho = \rho(r)$ or an equation 
of state $\rho = \rho(P)$ is prescribed. The most physical case is to 
prescribe an equation of state. Instead, in the third section 
solutions with constant density are investigated.  

Equation~(\ref{a}) can easily be integrated. 
Set the constant of integration equal to zero because of regularity at 
the centre.
Introduce the definitions of mass up to $r$ and mean density up to $r$ by
\begin{align}
	m(r)=4\pi\int_{0}^{r}s^2\rho(s)\dif s,\qquad
        w(r)=\frac{m(r)}{r^{3}},
	\label{md}
\end{align}
respectively. Then the first field equation~(\ref{a}) yields to
\begin{align}
      e^{-a(r)}=1-\lambda 2w(r)r^{2}-\lambda \frac{\Lambda}{3}r^{2}.
      \label{ea}
\end{align}
Therefore one may write metric~(\ref{sssmetric}) as
\begin{align}
      \dif s^{2}=-\frac{1}{\lambda}e^{\lambda \nu (r)}\dif t^{2}
      +\frac{\dif r^2}{1-\lambda 2 w(r)r^{2}
      -\lambda \frac{\Lambda}{3}r^{2}}
      +r^{2}\dif\Omega^{2}.
      \label{eqn_gen.metric}
\end{align}
The function $\nu'(r)$ can be eliminated from~(\ref{b}) and~(\ref{c}) 
and yields to the TOV-$\Lambda$ 
equation~\cite{Stuchlik:2000,Winter:2000ao}
\begin{align}
	P'(r)=-r\frac{\left( \lambda 4\pi P(r) + w(r) -\frac{\Lambda }{3} \right) 
		       \left( \lambda P(r)+ \rho(r) \right)}
		      {1-\lambda 2 w(r)r^{2}-\lambda \frac{\Lambda}{3}r^{2} },
	\label{tov}
\end{align}
which for $\Lambda =0$ gives the Tolman-Oppenheimer-Volkoff equation, 
without cosmological term, short TOV 
equation~\cite{Oppenheimer:1939ne,Tolman:1939}.

If an equation of state $\rho = \rho(P)$ is given~\cite{Rendall:1991hg}, 
the conservation equation~(\ref{c}) can 
be integrated to give
\begin{align}
	\nu(r)=- \int_{P_{c}}^{P(r)}\frac{2 \dif P}{\lambda P+\rho(P)},
	\label{int}
\end{align}
where $P_{c}$ denotes the central pressure. Using the definition of $m(r)$ then 
(\ref{md}) and (\ref{tov}) form an integro-differential system for $\rho(r)$ 
and $P(r)$. Differentiating (\ref{md}) with respect to $r$ implies
\begin{align}
	w'(r)=\frac{1}{r}\left(4\pi \rho(P(r)) - 3w(r) \right).
	\label{mdd}
\end{align}
Therefore given $\rho=\rho(P)$ equations (\ref{tov}) and (\ref{mdd}) are 
forming a system of differential equations in $P(r)$ and
$w(r)$. In~\cite{Rendall:1991hg} existence and uniqueness for a given
equation of state for this system was shown.
The solution for $w(r)$ together with~(\ref{ea}) gives the function $a(r)$, 
whereas the solution of $P(r)$ in (\ref{int}) defines $\nu(r)$.

\section{Solutions with constant density}
\label{secrho}

For practical reasons the notation is changed to geometrised units where $c^{2}=1/\lambda=1$.
Assume a positive constant density distribution $\rho =\rho_{0}=\text{const.}$ 
Then $w=\frac{4\pi}{3} \rho_{0}$ gives (\ref{tov}) in the form
\begin{align}
      P'(r)=-r\frac{\left(4\pi P(r)+\frac{4\pi}{3}\rho_{0}-\frac{\Lambda}{3}\right)
      \left( P(r) + \rho_{0} \right)}{1-kr^{2}},
\label{tov_const}
\end{align}
where $k$ is given by
\begin{align}
      k = \frac{8\pi}{3}\rho_{0} + \frac{\Lambda}{3}.
      \label{k}
\end{align}
In the following all solutions to the above differential equation are derived, see 
\cite{Stephani:1967,Stuchlik:2000,Tolman:1939}. 

The central pressure $P_{c}=P(r=0)$ is always assumed to be positive and finite.
Using that the density is constant in~(\ref{int}) metric~(\ref{eqn_gen.metric}) 
reads 
\begin{align}
      \dif s^{2}=-\left(\frac{P_{c}+\rho_{0}}{P(r)+\rho_{0}}\right)^{2}\dif t^{2}
      +\frac{\dif r^2}{1-k r^{2}}+r^{2}\dif \Omega^{2}.
      \label{eqn_metric}
\end{align}
For non-vanishing $k$ a new coordinate $\alpha$ will be introduced
\begin{align}
      k < 0,&\qquad r=\frac{1}{\sqrt{-k}}\sinh\alpha,
      \label{eqn_sinh} \\
      k > 0,&\qquad r=\frac{1}{\sqrt{k}}\sin\alpha.
      \label{eqn_sin}
\end{align}
The metric is well defined for radial coordinates $r \in \left[0,\hat{r}\right)$
if $\Lambda > -8\pi \rho_{0}$, where $\hat{r}$ denotes the zero of $g^{rr}$.
If $\Lambda \leq -8\pi \rho_{0}$ the metric is well defined for all $r$.

Solutions of differential equation (\ref{tov_const}) are uniquely determined
by the three parameters $\rho_{0}$, $P_{c}$ and $\Lambda$. Therefore
one has a 3-parameter family of solutions.

\subsection{Stellar models with spatially hyperbolic geometry} 
\subsection*{$\Lambda < -8 \pi \rho_{0}$}
If  $\Lambda < -8 \pi \rho_{0}$ then $k<0$ and the spatial geometry of~(\ref{eqn_metric}) 
is hyperbolic and the differential equation does not have a singularity. 
The volume of group orbits has no extrema, thus metric~(\ref{eqn_metric})
has no coordinate singularities and is well defined for all $r$.

Integration of~(\ref{tov_const}) yields to
\begin{align}
      P(\alpha)=\rho_{0}\frac{\left(1-\frac{\Lambda}{4\pi\rho_{0}}\right)-C \cosh\alpha}
      {C \cosh \alpha-3}.
      \label{pr<}
\end{align}
$C$ is the constant of integration, evaluated by defining \mbox{$P(\alpha =0)=P_{c}$} 
to be the central pressure. It is found that
\begin{align}
      C = \frac{3P_{c}+\rho_{0}-\frac{\Lambda}{4\pi}}{P_{c}+\rho_{0}}.
      \label{constantC}
\end{align} 
The function $P(\alpha)$ is well defined and monotonically decreasing for all $\alpha$.

The pressure converges to $-\rho_{0}$ as $\alpha$ or as the radius $r$ tends to infinity. 
Thus there always exists an $\alpha_b$ or $R$ where the pressure vanishes.
Vanishing of the numerator in~(\ref{pr<}) yields to
\begin{align}
        \cosh \alpha_{b} = \frac{1}{C} \left(1-\frac{\Lambda}{4 \pi \rho_{0}}\right),
        \label{eqn_coshalpha<}
\end{align}
the corresponding radius $R$ is given by
\begin{align}
      R^{2}=\frac{3}{|8 \pi \rho_{0}+\Lambda |}
      \left(\frac{1}{C^{2}}\left(1-\frac{\Lambda}{4 \pi \rho_{0}}\right)^{2} -1\right).
      \label{eqn_prpc<}
\end{align}
The new coordinate $\alpha$ simplifies expressions. Therefore the 
radial coordinate will only be used to give a physical picture or if it simplifies the 
calculations. 

From equation (\ref{eqn_coshalpha<}) one can deduce the inverse function. Then one has
the central pressure as a function of $\alpha_{b}$, $P_{c}=P_{c}(\alpha_{b})$. To do this the 
explicit expression for $C$ from (\ref{constantC}) is needed. One finds
\begin{align}
        P_{c}=\rho_{0} \frac{\left(1-\frac{\Lambda}{4 \pi \rho_{0}}\right) 
        \left(\cosh \alpha_{b} -1 \right)}
        {\left(1-\frac{\Lambda}{4 \pi \rho_{0}}\right) - 3\cosh \alpha_{b}}.
\label{eqn_pcpr<}
\end{align}

The central pressure given by (\ref{eqn_pcpr<}) should be finite. Therefore one obtains
an analogue of the Buchdahl~\cite{Buchdahl:1959} inequality. It reads
\begin{align}
        \cosh \alpha_{b} < \frac{1}{3}\left(1-\frac{\Lambda}{4 \pi \rho_{0}}\right).
\label{eqn_buchdahl<}
\end{align}
Thus there exists an upper bound for $\alpha_{b}$ for given $\Lambda$ and $\rho_{0}$.
Use that $\sinh(\arccosh (\alpha)) = \sqrt{\alpha^{2}-1}$,
then the corresponding radius $R$ is given by
\begin{align}
      R^{2} < \frac{\frac{1}{3} \left(4-\frac{\Lambda}{4 \pi \rho_{0}} \right)}{4 \pi \rho_{0}}.
      \label{eqn_R<}
\end{align}
Since the cosmological constant is negative the right-hand side 
of (\ref{eqn_R<}) is well defined. Using the definition of 
mass $M=(4\pi/3)\rho_{0} R^{3}$ one can rewrite (\ref{eqn_R<})
in terms of $M$, $R$ and $\Lambda$. This leads to
\begin{align}
	3M < \frac{2}{3}R + R\sqrt{\frac{4}{9}-\frac{\Lambda}{3} R^{2}}.
	\label{eqn_3m<_hyp}
\end{align}
This is the wanted analogue of the Buchdahl inequality (\ref{eqn_buchdahl}).

At $r=R$, where the pressure vanishes, the Schwarzschild-anti-de Sitter 
solution is joined.

\subsection*{Joining interior and exterior solution} 
At the $P=0$ surface the Schwarzschild-anti-de Sitter solution 
is joined by using Gauss coordinates to the $r=\mbox{const.}$ hypersurfaces
and defining the mass by $M=(4\pi/3)\rho_{0} R^{3}$.
If necessary one can rescale the time to get the $g_{tt}$-component 
continuous. With this rescaling and the use of Gauss coordinates
one can get metric $C^{1}$ at the boundary.

Since the density is not continuous at the boundary the Ricci tensor is not, either. 
Therefore the metric is at most $C^{1}$. This cannot be improved.

\subsection{Stellar models with spatially Euclidean geometry}
\subsection*{$\Lambda =-8 \pi \rho_{0}$}
Assume that cosmological constant and constant density are chosen such that
$8\pi \rho_{0} = -\Lambda$.  Then $k=0$ and the denominator of 
(\ref{tov_const}) becomes one and the differential equation simplifies to
\begin{align}
      \frac{\dif P}{\dif r}=-4\pi r\left( P+\rho_{0} \right)^{2},
      \label{tov_special}
\end{align}
and the $t=\text{const.}$ hypersurfaces of (\ref{eqn_metric}) are purely Euclidean.
As in the former case metric (\ref{eqn_metric}) is well defined for all $r$.
Integration yields to 
\begin{align}
      P(r)=\frac{1}{2\pi r^{2}+\frac{1}{P_{c}+ \rho_{0}}} - \rho_{0},
      \label{pr_special}
\end{align}
where the constant of integration is fixed by $P(r=0)=P_{c}$.

The denominator of the pressure distribution cannot vanish because
central pressure and density are assumed to be positive. 
Therefore (\ref{pr_special}) has no singularities.
As the radius tends to infinity the fraction tends to zero and
thus the pressure converges to $-\rho_{0}$.

This implies that there always exits a radius $R$ where $P(r=R)=0$.
Therefore all solutions to (\ref{tov_special}) in (\ref{eqn_metric}) 
describe stellar objects. Their radius is given by 
\begin{align}
	R^{2}=\frac{1}{2\pi} \left( \frac{1}{\rho_{0}} - \frac{1}{P_{c}+\rho_{0}} \right).
	\label{eqn_prpc=}
\end{align}

One finds that the radius $R$ is bounded by $1/\sqrt{2\pi \rho_{0}}$.
Inserting the definition of the density yields to
\begin{align}
	R^{2} < \frac{1}{2 \pi \rho_{0}} = \frac{1}{2 \pi} \frac{4 \pi R^{3}}{3M},
\label{eqn_R<1/2pilambda}
\end{align}
which implies 
\begin{align}
	3 M < 2 R. 
\label{eqn_M<23R}
\end{align}
Thus (\ref{eqn_M<23R}) is the analogous Buchdahl inequality to
equation~(\ref{eqn_buchdahl}) and equals (\ref{eqn_3m<_hyp}) with
$\Lambda=-8\pi \rho_{0}=-6M/R^{3}$.

At $r=R$, where the pressure vanishes, the Schwarzschild-anti-de Sitter
solution is joined uniquely as described.
Since the density is not continuous at the boundary the Ricci tensor is not, either. 
Thus the metric is again at most $C^{1}$.

\subsection{Stellar models with spatially spherical geometry}
\label{lambda>-8pirho}
\subsubsection*{$\Lambda > -8 \pi \rho_{0}$}
If $\Lambda > -8 \pi \rho_{0}$ then $k$ is positive.
Equation (\ref{tov_const}) has a singularity at
\begin{align}
        r = \hat{r} = \frac{1}{\sqrt{k}}.
\label{eqn_def_hatr}
\end{align}
In $\alpha$ the singularity $\hat{r}$ corresponds to $\alpha(\hat{r}) = \pi /2$,
the equator of a 3-sphere.
The spatial part of metric~(\ref{eqn_metric}) now describes a part of the 3-sphere 
of radius $1/\sqrt{k}$. The metric is well defined for radii less than $\hat{r}$.

Integration of~(\ref{tov_const}) gives
\begin{align}
      P(\alpha)=\rho_{0}\frac{\left(\frac{\Lambda}{4 \pi \rho_{0}}-1\right)
      +C\cos \alpha}{3-C\cos \alpha},
      \label{eqn_pr<>}
\end{align}
where the constant of integration $C$ is given by~(\ref{constantC}).
One finds 
\begin{align}
        P(\pi /2)=\frac{\rho_{0}}{3} \left(\frac{\Lambda}{4 \pi \rho_{0}}-1\right),
\label{eqn_phatr}
\end{align}
which is less than zero if $\Lambda < 4 \pi \rho_{0}$,
which is the restriction in~\cite{Stuchlik:2000} and the
output of~\cite{Nowakowski:2000dr,Nowakowski:2001zw}.  
Since this is the considered case the singularity 
of the pressure gradient is not important yet.

Since $P(\pi /2) < 0$ there exists an $\alpha_{b}$ such that $P(\alpha_{b})=0$.
$\alpha_{b}$ can be derived from equation (\ref{eqn_pr<>}) and one obtains
the analogue of (\ref{eqn_prpc<})
\begin{align}
       \cos \alpha_{b} = \frac{1}{C} \left( 1-\frac{\Lambda}{4 \pi \rho_{0}} \right).
\label{eqn_prpc<>}
\end{align}

It remains to derive the analogue of the Buchdahl inequality for this case. 
One uses (\ref{eqn_prpc<>}) to find $P_{c}=P_{c}(\alpha_{b})$
\begin{align}
      P_{c}=\rho_{0} \frac{\left(1-\frac{\Lambda}{4 \pi \rho_{0}}\right) 
      \left(1-\cos \alpha_{b} \right)}
      {3\cos \alpha_{b} - \left(1-\frac{\Lambda}{4 \pi \rho_{0}}\right)},  
      \label{eqn_pcpr<>}
\end{align}
which is similar to (\ref{eqn_prpc<}).
Finiteness of the central pressure in (\ref{eqn_pcpr<>}) gives 
\begin{align}
        \cos \alpha_{b} > \frac{1}{3}\left(1-\frac{\Lambda}{4 \pi \rho_{0}}\right).
        \label{eqn_Buch}
\end{align}
With $\sin(\arccos(\alpha)) = \sqrt{1-\alpha^{2}}$,
this yields to the wanted analogue
\begin{align}
        R^{2} < \frac{\frac{1}{3} \left(4-\frac{\Lambda}{4 \pi \rho_{0}} \right)}{4 \pi \rho_{0}},
\label{eqn_buchdahl<>}
\end{align}
which equals (\ref{eqn_R<}). The greater sign reversed because $(\arccos\alpha)$ is decreasing.
Thus it can also be rewritten to give~(\ref{eqn_3m<_hyp}). 
At $\alpha=\alpha_{b}$ or $r=R$ the Schwarzschild-anti-de Sitter solution 
is joined and the metric is $C^{1}$. Without further assumptions this cannot be improved.

\subsection{Solutions with vanishing cosmological constant}
\subsection*{$\Lambda =0$}
Assume a vanishing cosmological constant. Then one can use all equations of the
former case with $\Lambda=0$. Only one of these relations will be shown, namely the
Buchdahl inequality \cite{Buchdahl:1959}. It is derived from (\ref{eqn_buchdahl<>})
\begin{align}
	R^{2} < \frac{1}{3\pi \rho_{0}},
\end{align}
using $M=(4\pi/3)\rho_{0} R^{3}$ leads to
\begin{align}
	M < \frac{4}{9} R,\qquad 3M < \frac{4}{3} R.
	\label{eqn_buchdahl}
\end{align} 

\subsection{Stellar models with spatially spherical geometry}
\subsection*{$0<\Lambda < 4 \pi \rho_{0}$}
Integration of the TOV-$\Lambda$ equation gives~(\ref{eqn_pr<>}).
Because of~(\ref{eqn_phatr}) the boundary $P(R)=0$ exists.  
As in the former sections one finds~(\ref{eqn_R<})
and written in terms of $M$, $R$ and $\Lambda$ again gives
\begin{align}
	3M < \frac{2}{3}R + R\sqrt{\frac{4}{9}-\frac{\Lambda}{3} R^{2}}.
	\label{eqn_3m<_lpos}
\end{align}
Since the cosmological constant is positive the square root term is well
defined if 
\begin{align}
	R \leq \sqrt{\frac{4}{3}} \frac{1}{\sqrt{\Lambda}}.
\end{align}
In this section $0<\Lambda < 4 \pi \rho_{0}$ is assumed. Using the definition
of mass this can be rewritten to give
\begin{align}
      R^2\Lambda < 4\pi \rho_{0}R^2 = \frac{3M}{R} 
      <\frac{2}{3}+\sqrt{\frac{4}{9}-\frac{\Lambda}{3}R^{2}},
\end{align}
where the right-hand side of~(\ref{eqn_3m<_lpos}) was used. 
By taking the $2/3$ on the other side and squaring the equation
this reduces to the simple inequality
\begin{align}
      R^2 \Lambda < 1,\qquad R<\frac{1}{\sqrt{\Lambda}},
      \label{eqn_bound_radius}
\end{align}
which means that the boundary of the stellar object is located 
before the cosmological event horizon.

It is remarkable that bounds on mass and radius similar 
to~(\ref{eqn_3m<_lpos}) and~(\ref{eqn_bound_radius}) were derived 
in~\cite{Nowakowski:2000dr,Nowakowski:2001zw} by considering the
virial theorem in the presence of a positive cosmological constant.
Furthermore the inequality $\Lambda<4\pi\rho$ is derived, which in
the context of the this paper may be seen as an evidence that
the exterior spacetime of stellar objects is the Schwarzschild-de Sitter
spacetime with increasing group orbits at the boundary. 

As before the Schwarzschild-de Sitter solution is joined $C^{1}$.
Since the density is not continuous at the boundary this cannot be improved. 

\begin{figure}[ht!]
\begin{center}
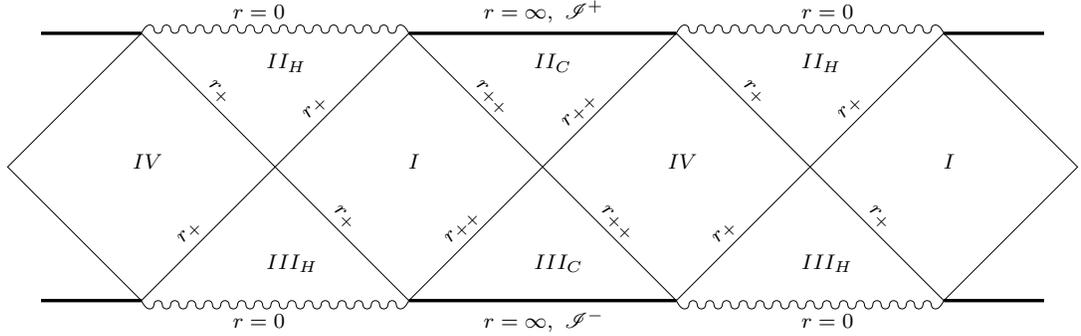
\end{center}
\caption{Penrose-Carter diagram for Schwarzschild-de Sitter space. The Killing
	vector $\partial/\partial t$ is time-like and future-directed in regions $I$ and
	time-like and past-directed in regions $IV$. In the others regions it is space-like.
	The surfaces $r=r_{+}$ and $r=r_{++}$ are black-hole and cosmological event
	horizons, respectively. $\scri^{+}$ and $\scri^{-}$ are the space-like infinities.}
\label{fig: Penrose}
\end{figure}

\begin{figure}[ht!]
\begin{center}
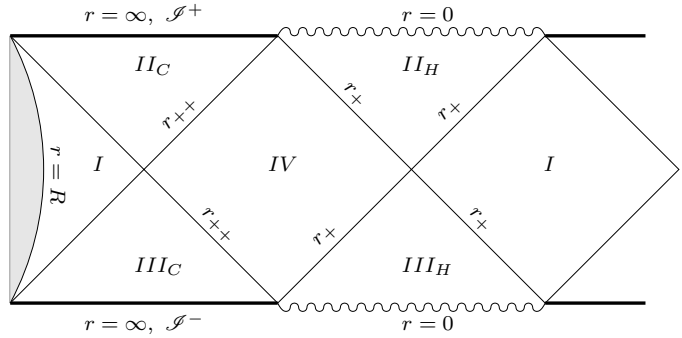
\end{center}
\caption{Penrose-Carter diagram with one stellar object having a radius $R$ which lies between
	the two horizons. The group orbits are increasing at the boundary. 
	The shaded region is the matter solution with regular centre. 
	There is still an infinite sequence of  singularities $r=0$ and 
	space-like infinities $r=\infty$. }
\label{fig: Penrose1a}
\end{figure}

\begin{figure}[ht!]
\begin{center}
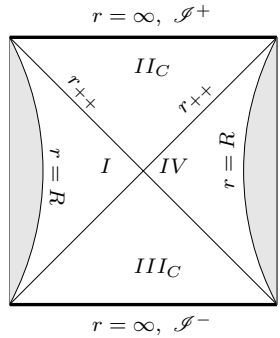
\end{center}
\caption{Penrose-Carter diagram with two stellar objects having radii $R$ which lie between
	the two horizons. Since the group orbits are increasing up to $R$ the vacuum 
	part contains the cosmological event horizon $r_{++}$. 
	This solution with two objects has no singularities. Because of regions $II_{C}$ 
	and $III_{C}$ this spacetime is not globally static.}
\label{fig: Penrose2}
\end{figure}

One can construct solutions without singularities.
The group orbits are increasing up to $R$. 
The boundary of the stellar object $r=R$ can be put in region $I$ of 
Penrose-Carter figure~\ref{fig: Penrose} where the time-like 
killing vector is future directed. This leads to figure~\ref{fig: Penrose1a},
the spacetime still contains an infinite sequence of singularities $r=0$
and space-like infinities $r=\infty$.
It is possible to put a second object with boundary $r=R$ in region $IV$ of Penrose-Carter
figure~\ref{fig: Penrose1a} where the time-like killing vector is past directed.
This leads to figure~\ref{fig: Penrose2} and shows that there 
are no singularities in the spacetime. This spacetime is not globally
static because of the remaining dynamical parts of the the
Schwarzschild-de Sitter spacetime, regions $II_{C}$ and $III_{C}$.
Penrose-Carter diagrams are in the appendix~\ref{app_penrose}.

A black hole alternative solution without singularities and without horizons was
described by~\cite{Mazur:2001fv}.

\subsection{Solutions with exterior Nariai metric}
\label{subsec_ext_nariai}
\subsection*{$\Lambda = 4 \pi \rho_{0}$}
In this special case integration of (\ref{tov_const}) gives
\begin{align}
        P(\alpha) = \rho_{0} \frac{C\cos \alpha}{3-C\cos \alpha}.
\label{eqn_pr_4pi}
\end{align}
The pressure (\ref{eqn_pr_4pi}) vanishes at $\alpha = \alpha_{b} = \pi /2$, 
the coordinate singularity in the radial coordinate $r$, the interior metric reads
\begin{align}
      \dif s^{2}= -\left(1-\frac{P_{c}}{P_{c} + \rho_{0}}\cos \alpha\right)^{2} \left(\frac{P_{c}+\rho_{0}}
      {\rho_{0}}\right)^{2} \dif t^{2} 
      +\frac{1}{\Lambda}\left[\dif\alpha^{2}+\dif\Omega^{2}\right].
      \label{eqn_interior_4pi}
\end{align}
One would like to join the Schwarzschild-de Sitter metric, but this does not work,
since the group orbits of the Schwarzschild-de Sitter metric are increasing whereas
the group orbits of the interior metric have constant volume.
But there is one other spherically symmetric vacuum solution to 
the Einstein field equations with cosmological constant, the 
Nariai solution \cite{Nariai:1999nc,Nariai:1999os}. Its metric is 
\begin{align}
        \dif s^{2}=\frac{1}{\Lambda}
        \left[-(A \cos \log r + B \sin \log r)^{2} \dif t^{2} 
        + \frac{1}{r^{2}} \dif r^{2}  + \dif \Omega^{2} \right],
\label{eqn_nariai_r}
\end{align}
where $A$ and $B$ are arbitrary constants. With $r=e^{\alpha}$ this becomes
\begin{align}
      \dif s^{2}=\frac{1}{\Lambda} 
      \left[-(A \cos \alpha + B\sin\alpha)^{2}\dif t^{2} 
      +\dif\alpha^{2}+\dif\Omega^{2}\right].
      \label{eqn_nariai_a}
\end{align}
Metrics (\ref{eqn_interior_4pi}) and (\ref{eqn_nariai_a}) can be joined by fixing the 
constants $A$ and $B$. With
\begin{align}
      A=-\frac{P_{c}}{\rho_{0}}\sqrt{\Lambda},\qquad
      B=\frac{P_{c}+\rho_{0}}{\rho_{0}}\sqrt{\Lambda},
\end{align}
the metric is $C^{1}$ at $\alpha=\pi /2$. As before, since the density is not continuous, the metric
is at most $C^{1}$.

\subsection{Solutions which have decreasing group orbits at the boundary}
\label{subsec_ext_sing}
\subsection*{$4 \pi \rho_{0} < \Lambda < \Lambda_{0}$}
Integration of the TOV-$\Lambda$ equation again yields~(\ref{eqn_pr<>}).
Assume that the pressure vanishes before the second centre $\alpha=\pi$ of
the 3-sphere is reached. The condition $P(\alpha=\pi)<0$ leads to an
upper bound for the cosmological constant. This bound is given by
\begin{align}
        \Lambda_{0} := 4 \pi \rho_{0} \left( \frac{4 \, P_{c}/\rho_{0} +2} {P_{c}/\rho_{0}+2} \right).
\end{align}
Then $4 \pi \rho_{0} < \Lambda < \Lambda_{0}$ implies the following:

The pressure is decreasing near the centre and vanishes for some $\alpha_{b}$, where 
\mbox{$\pi /2 < \alpha_{b} < \pi$}. Equation~(\ref{eqn_sin}) together with the 
metric~(\ref{eqn_metric}) implies that the volume of the group orbits is decreasing 
if $\alpha > \pi /2$. 

\begin{figure}[ht!]
\begin{center}
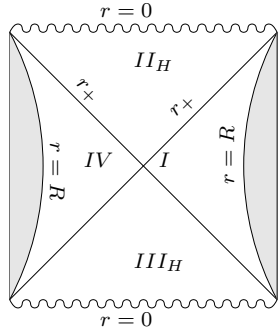
\end{center}
\caption{Penrose-Carter diagram with two stellar objects having radii $R$ which lie between
	the two horizons. The group orbits of the interior solutions are decreasing where
	the vacuum solution is joined. Thus the $r=0$ singularity of the vacuum part is present.}
\label{fig: Penrose3}
\end{figure}

At $\alpha_{b}$ one uniquely joins the Schwarzschild-de Sitter solution by $M=(4\pi/3)\rho_{0}R^{3}$. 
With Gauss coordinates relative to the $P(\alpha_{b})=0$ hypersurface the metric will 
be $C^{1}$. But there is a crucial difference to the former case with exterior 
Schwarzschild-de Sitter solution. Because of the decreasing group orbits at the 
boundary there is still the singularity $r=0$ in the
vacuum spacetime. Penrose-Carter diagram~\ref{fig: Penrose3} shows this interesting solution.  

\subsection{Decreasing solutions with two regular centres}
\label{subsec_2centres_a}
\subsection*{$\Lambda_{0} \leq \Lambda < \Lambda_{E}$}
As before the pressure is given by~(\ref{eqn_pr<>}).  
Assume that the pressure is decreasing near the first centre $\alpha=0$.
This gives an upper bound of the cosmological constant
\begin{align}
	\Lambda_{E} := 4 \pi \rho_{0} \left( 3 \, P_{c}/\rho_{0} +1 \right),
\label{lambdaE}
\end{align}
where $\Lambda_{E}$ is the cosmological constant of the Einstein static universe.
These possible values $\Lambda_{0} \leq \Lambda < \Lambda_{E}$ imply:

The pressure is decreasing is near the first centre $\alpha=0$ but remains
positive for all $\alpha$ because $\Lambda \geq \Lambda_{0}$. 
Therefore there exists a second centre at $\alpha=\pi$. At the second
centre of the 3-sphere the pressure~(\ref{eqn_pr<>}) becomes
\begin{align}
	P(\alpha=\pi) = \rho_{0} \frac{ \left( \frac{\Lambda}{4 \pi \rho_{0}} -1 \right) -C}{3 + C}.
	\label{eqn_p_sec_centre}
\end{align}
It only vanishes if $\Lambda = \Lambda_{0}$. The solution is inextendible. 
The second centre is also regular. This is easily shown with Gauss
coordinates.

Solutions of this kind are generalisations of the Einstein static universe. These 3-spheres
have a homogeneous density but do not have constant pressure. They have a given central
pressure $P_{c}$ at the first regular centre which decreases monotonically towards the second 
regular centre. Generalisations of the Einstein static universe were also found earlier
in Ref.~\cite{Ibrahim:1976} by Ibrahim and Nutku.\footnote{I would like to thank Aysel 
Karafistan for bringing this reference to my attention.}

\subsection{The Einstein static universe}
\label{subsec_einstein}
\subsection*{$\Lambda=\Lambda_{E}$}
Assume a constant pressure function. Then the pressure gradient vanishes and
therefore the right-hand side of~(\ref{tov_const}) has to vanish. This gives 
an unique relation between density, central pressure and cosmological constant.
One obtains~(\ref{lambdaE})
\begin{align}
        \Lambda = \Lambda_{E} = 4\pi \left( 3P_{E} + \rho_{0} \right),
\end{align}
where $P_{E}=P_{c}$ was used to emphasise that the given central pressure corresponds
to the Einstein static universe~\cite{Einstein:1917ce} and is homogeneous. 
Its metric is of no further interest.

For a given density $\rho_{0}$ there exists for every choice of a central pressure 
$P_{c}$ a unique cosmological constant given by~(\ref{lambdaE}) such that an Einstein 
static universe is the solution to~(\ref{tov_const}).

\subsection{Increasing solutions with two regular centres}
\subsection*{$\Lambda_{E} < \Lambda < \Lambda_{S}$}
Assume that the pressure~(\ref{eqn_pr<>}) is finite at the second centre.
This leads to an upper bound of the cosmological constant defined by
\begin{align}
        \Lambda_{S} := 4\pi \rho_{0} \left( 6 \, P_{c}/\rho_{0} + 4 \right).
\end{align}
The possible values of the cosmological constant imply:

The pressure $P(\alpha)$ is increasing near the first regular centre. It
increases monotonically up to $\alpha = \pi$, where one has a second regular centre.
This situation is similar to the case where  
\mbox{$\Lambda_{0} \leq \Lambda < \Lambda_{E}$}.
These solutions are also describing generalisations of the 
Einstein static universe.

The generalisations are symmetric with respect to the Einstein 
static universe. By symmetric one means the following. 
Instead of writing the pressure as a function of $\alpha$
depending on the given values $\rho_{0}$, $P_{c}$ and $\Lambda$ 
one can eliminate the cosmological constant with the pressure at 
the second centre $P(\alpha=\pi)$, given by (\ref{eqn_p_sec_centre}). 
Then one easily finds that the pressure is symmetric to $\alpha=\pi/2$ 
if both central pressures are exchanged and therefore this is the
converse situation to the case where $\Lambda_{0} \leq \Lambda < \Lambda_{E}$.
If the central pressures are equal the dependence on $\alpha$ 
vanishes and one is left with the Einstein static universe.

\subsection{Solutions with geometric singularity}
\label{subsec_sing}
\subsection*{$\Lambda \geq \Lambda_{S} $}
In this case it is assumed that $\Lambda$ exceeds the upper limit $\Lambda_{S}$.
Then~(\ref{eqn_pr<>}) implies that the pressure is increasing near the centre and
diverges before $\alpha = \pi$ is reached. Thus these solutions have a geometric 
singularity with unphysical properties, as can be seen by considering the square of
the Riemann tensor. Therefore they are of no further interest.  

\subsection*{Acknowledgements}
I would like to thank Bernd G. Schmidt who supervised my diploma 
thesis~\cite{Boehmer:2002gg} on which this paper is based.

Furthermore I am grateful to Herbert Balasin, Daniel Grumiller
and Wolfgang Kummer for their support and Marek Nowakowski for
pointing out the similarities between our works and for suggestions
concerning the manuscript.
\newpage
\appendix
\section{Overview of constant density solutions}
\label{app_over}
\begin{table}[ht!]
\begin{tabular}[t]{|c|c|p{.55\linewidth}|} \hline
cosmological & spatial & short description of the solution\\ 
constant & geometry & \mbox{} \\ \hline
$\Lambda<-8\pi \rho_{0}$ 		& hyperboloid 	& stellar model with exterior Schwarzschild-anti-de Sitter solution \\
\mbox{} 				& \mbox{} 	& \mbox{} \\
$\Lambda=-8\pi \rho_{0}$		& Euclidean	& stellar model with exterior Schwarzschild-anti-de Sitter solution \\
\mbox{} 				& \mbox{} 	& \mbox{} \\
$-8\pi \rho_{0}<\Lambda<0$		& 3--sphere	& stellar model with exterior Schwarzschild-anti-de Sitter solution  \\ 
\mbox{} 				& \mbox{} 	& \mbox{} \\ \hline
$\Lambda =0 $			& 3--sphere	& stellar model with exterior Schwarzschild solution  \\ 
\mbox{} 				& \mbox{} 	& \mbox{} \\ \hline
$0<\Lambda<4\pi \rho_{0}$		& 3--sphere	& stellar model with exterior Schwarzschild-de Sitter solution  \\
\mbox{} 				& \mbox{} 	& \mbox{} \\
$\Lambda=4\pi \rho_{0}$		& 3--sphere	& stellar model with exterior Nariai solution  \\
\mbox{} 				& \mbox{} 	& \mbox{} \\
$4\pi \rho_{0}<\Lambda<\Lambda_{0} $	& 3--sphere	& decreasing pressure model with exterior Schwarzschild-de Sitter solution; the group orbits are decreasing at the boundary			\\
\mbox{} 				& \mbox{} 	& \mbox{} \\
$\Lambda_{0}\leq\Lambda<\Lambda_{E}$	& 3--sphere	& solution with two centres, pressure decreasing near the first; generalisation of the Einstein static universe 			\\	
\mbox{} 				& \mbox{} 	& \mbox{} \\
$\Lambda=\Lambda_{E}$		& 3--sphere	& Einstein static universe			\\
\mbox{} 				& \mbox{} 	& \mbox{} \\
$\Lambda_{E}<\Lambda<\Lambda_{S}$	& 3--sphere	& solution with two centres, solution increasing near the first; generalisation of the Einstein static universe 	\\
\mbox{} 				& \mbox{} 	& \mbox{} \\
$\Lambda \geq \Lambda_{S}$		& 3--sphere	& increasing pressure solution with geometric singularity \\ \hline
\end{tabular}
\caption{Overview of constant density solutions}
\label{oview}
\end{table}

\section{Penrose-Carter diagrams}
\label{app_penrose}

\begin{figure}[ht!]
\begin{center}
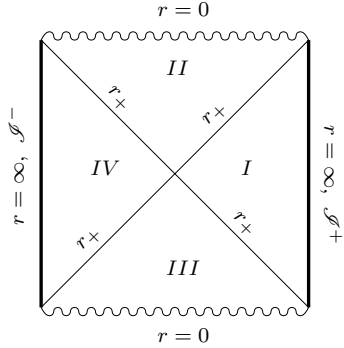
\end{center}
\caption{Penrose-Carter diagram for Schwarzschild-anti-de Sitter space. 
	The surface $r=r_{+}$ is the black-hole event horizons.
	$\scri^{+}$ and $\scri^{-}$ are the time-like infinities.}
\label{fig: Penrose_anti1}
\end{figure}

\newpage \mbox{} \newpage

\addcontentsline{toc}{section}{References}
\bibliographystyle{plain}
\bibliography{review}

\end{document}